\documentclass[useAMS,usenatbib]{mn2e}
\usepackage{natbib}
\usepackage{graphicx}
\usepackage{listings}
\usepackage[intlimits]{amsmath}
\usepackage{txfonts}
\usepackage{bm}
\usepackage{amssymb}
\usepackage{amsmath}
\usepackage{tabularx} 
\usepackage{color}

\usepackage{comment}

\title[Current closure through the neutron star crust]{Current closure through the neutron star crust}

\author[V. Karageorgopoulos, K.N. Gourgouliatos \& I. Contopoulos]{{V. Karageorgopoulos$^{1,2}$\thanks{Email: vkarageo@upatras.gr}, K. N. Gourgouliatos\thanks{Email: Konstantinos.Gourgouliatos@durham.ac.uk}$^{2}$ \& I. Contopoulos$^{3}$\thanks{Email: icontop@academyofathens.gr}}\vspace{0.4cm}\\
\parbox{\textwidth}{$^{1}$ University of Patras, Department of Physics,
  Patras Greece
}\\
\parbox{\textwidth}{$^{2}$Department of Mathematical Sciences, Durham University, Durham DH1 3LE, UK }\\
\parbox{\textwidth}{$^{3}$Research Centre for Astronomy and Applied Mathematics, Academy of Athens,
 Athens, Greece} }

\begin{document}

\date{Accepted -. Received -; in original form -}
\pagerange{\pageref{firstpage}--\pageref{lastpage}} \pubyear{-}

\maketitle

\label{firstpage}

\begin{abstract}
Force-free pulsar magnetospheres develop a large scale poloidal electric current circuit that flows along open magnetic field lines from the neutron star to the termination shock. The electric current closes through the interior of the neutron star where it provides the torque that spins-down the star. In the present work, we study the internal electric current in an axisymmetric rotator. We evaluate the path of the electric current by requiring the minimization of internal Ohmic losses.
We find that, in millisecond pulsars, the current reaches the base of the crust, while in pulsars with periods of a few seconds, the bulk of the electric current does not penetrate deeper than about $100$~m. The region of maximum spin-down torque in millisecond pulsars is the base of the crust, while in slowly spinning ones it is the outer crust. We evaluate the corresponding Maxwell stresses and find that, in typical rotation-powered radio pulsars, they are well below the critical stress that can be sustained by the crust. For magnetar-level fields, the Maxwell stresses near the surface are comparable to the critical stress and may lead to the decoupling of the crust from the rest of the stellar rotation.
\end{abstract}

\maketitle


\section{Introduction}
\label{intro}

A rotating magnetized neutron star is surrounded by a plasma-filled electrically conducting force-free magnetosphere in which the magnetic field is energetically dominant and governs its overall dynamics \citep{Goldreich:1969}. Time-dependent electrodynamic and magnetohydrodynamic numerical simulations relax to a steady-state ideal force-free magnetic field configuration that corotates with the neutron star \citep{Spitkovsky:2006, Komissarov:2006, Tchekhovskoy:2013}.   This magnetospheric solution was first obtained by \cite{Contopoulos:1999} in the case of an axisymmetric rotator.

An important characteristic of the steady-state configuration is that magnetic field lines that cross the light-cylinder contain a certain distribution of poloidal electric current that forms a large scale electric circuit. This is associated with a toroidal magnetic field component that reflects the fact that magnetic field lines are swept backwards with respect to the stellar rotation. Without it, plasma `frozen into' these field lines beyond the light cylinder would move faster than the speed of light. This electric current distribution is the only one that guarantees smooth crossing of the light cylinder by the magnetic field, and  in that sense, it is an `eigenfunction' of the problem. 

The `generator' (or `battery') of the magnetospheric electric circuit is the neutron star rotation, the `wires' are the magnetic flux surfaces, and the `loads' are finite dissipation regions at large distances (near and beyond the light cylinder, and the termination shock at very large distances). The electric current closes through the stellar interior. It penetrates deep inside the crust where it generates the torques necessary to spin down the neutron star (force-free conditions must be abandoned there). 

The crust comprises an exceptionally strong ion lattice, nevertheless, it can only sustain finite stresses \citep{Strohmayer:1991, Chamel:2008, Horowitz:2015}. If the spin-down torque is exerted on a very thin volume, the Maxwell stresses could in principle exceed the yield limit of the crust, and the crust would yield. While the magnetospheric solution is obtained by assuming an ideal plasma with infinite conductivity, the crust has a high but finite conductivity $\sigma$ ranging between $10^{20}$ and $10^{27}$~s$^{-1}$ \citep{Potekhin:2015}. We note that even if there is a finite resistivity in the magnetosphere \citep{Li:2012, Kalapotharakos:2012}, the big picture does not change qualitatively. 

In this paper we obtain the flow of the magnetospheric electric current inside the neutron star crust and calculate the transfer of magnetospheric spindown torque into the stellar interior. The plan of the paper is as follows. In section~2, we derive the governing equations that describe the flow of electric current in the stellar interior. In section~3, we solve these equations numerically and present results for several pulsar models. We discuss their implications in section~4, and present our conclusions in section~5.  

\section{Problem setup}

\subsection{The equation for the electric current in the crust}
\label{ohlm}

Let us consider an axisymmetric stationary configuration. In what follows, we will work in spherical coordinates $(r,\theta,\phi)$ centered on the neutron star and aligned with the axis of symmetry (which coincides with the axis of rotation and the magnetic axis). The electric current density may in general be expressed as
\begin{eqnarray}
{\bf j}=\frac{1}{2\pi}  \nabla
I \times \nabla \phi +
j_\phi\hat{\phi}\ ,
\label{CURRENT}
\end{eqnarray}
where $I=I(r,\theta)$ is the electric current that passes through a ring perpendicular to and concentric with the axis of symmetry passing through position $(r,\theta)$. Notice that $I$ is related to the toroidal component of the magnetic field ${\bf B}$ as
\begin{eqnarray}
I=\frac{c}{2}r \sin\theta\ B_{\phi}\ ,
\label{Bfeq}
\end{eqnarray}
and we further assume that the poloidal component of the magnetic field is a dipole. 
The azimuthal component of the electric current $j_\phi$ in eq.~(\ref{CURRENT}) is due to the corrotation of the internal space-charge, namely $j_\phi=r~\sin\theta~ \Omega ~\nabla\cdot{\bf E}$, and does not enter in our calculations below. $\Omega$ is the stellar angular velocity and the electric field is given by Ohm's law
\begin{eqnarray}
{\bf E}=-r~\sin\theta~ \Omega ~{\bf{\bm{\hat{\phi}}}}\times {\bf B}/c+\frac{{\bf j}}\sigma\,.
\label{ELECTRIC}
\end{eqnarray}
Here $c$ is the speed of light and $\sigma$ the electric conductivity of the crust. The power per unit volume in the crust is given by: 
\begin{eqnarray}
{\bf j}\cdot {\bf E}= -r~\sin\theta~ \Omega ~\left({\bf{\bm{\hat{\phi}}}}\times {\bf B}/c\right)\cdot {\bf j}+\frac{{\bf j^2}}\sigma=\frac{1}{c} \left({\bf j}\times {\bf B}\right)\cdot {\bf v}+\frac{j^2}\sigma,
\label{POWER}
\end{eqnarray}
where ${\bf v}= r~\sin\theta ~\Omega~ {\bf{\bm{\hat{\phi}}}}$ is the velocity of the crust at $(r,\theta)$ for an observer in the lab frame. The term $\left({\bf j}\times {\bf B}\right)\cdot {\bf v}/c$ in eq.~(\ref{POWER}) expresses the work per unit volume and time done by the Lorentz force that spins-down the pulsar. The $j^2/\sigma$ term is the Ohmic thermal losses per unit volume and time, due to the finite conductivity of the crust. 

Our goal is to obtain the distribution $I(r,\theta)$ in the stellar interior. We will approach this question by applying a Fermat-type principle. We propose that the current inside the crust of the neutron star will distribute itself so that it minimises the total Ohmic thermal losses. This allows us to formulate a minimisation equation.
\begin{eqnarray}
P_{\rm Ohm}\equiv 
\int_V \frac{j^2}{\sigma}\ {\rm d}V=\left(\frac{c}{4\pi}\right)^2 \int_V \frac{\left(\nabla \times {\bf B}\right)^2}{\sigma}\ {\rm d}V\ .
\label{OHMIC_POWER}
\end{eqnarray}
where we have used that ${\bf j}=(c/4\pi)\nabla \times {\bf B} $.  
By demanding that $P_{\rm Ohm}$ is minimized, we obtain the condition
\begin{eqnarray}
\frac{c}{4\pi} \nabla \times \left(\frac{\nabla \times {\bf B}}{\sigma}\right)\equiv \frac{c}{4\pi} \nabla \times \left(\frac{\bf j}{\sigma}\right)=0
\label{MINIM}
\end{eqnarray}
(see Appendix~\ref{Appendix1}). The same result is obtained if we start from Ohm's law in the stellar interior (eq.~\ref{ELECTRIC})
and realise that, for a stationary configuration, $\nabla \times {\bf E}={\bf 0}$ (Faraday's law), this yields
\begin{eqnarray}
\nabla \times \left(\frac{\bf j}{\sigma}\right)={\bf 0}\,, 
\end{eqnarray}
which is identical to eq.~(\ref{MINIM}). With the help of eq.~(\ref{CURRENT}), eq.~(\ref{MINIM}) then takes the form
\begin{eqnarray}
\frac{\partial^2 I}{\partial r^2} - \frac{1}{\sigma} \frac{\partial\sigma}{\partial r} \frac{\partial I}{\partial r} - \frac{\cos \theta}{ r^2\sin \theta} \frac{\partial I}{\partial \theta} + \frac{1}{r^2}\frac{\partial^2 I }{\partial \theta^2}=0 
\label{MINIMdeq}
\end{eqnarray}
where we assume that the electric conductivity of the crust is a function of radius only $\sigma=\sigma(r)$. 

The two approaches are interconnected. Equation \ref{MINIM} essentially describes an Ohmic eigenmode \citep{Chanmugam:1972} corresponding to the zero eigenvalue, or equivalently infinite decay time, subject to given boundary conditions. The infinite decay time is imposed here by setting $\nabla \times {\bf E}=0$. The Ohmic thermal power minimisation derivation, starts from Ohm's law as well. Here we assume that among all possible electric current configurations that are compatible with the boundary conditions the one that will survive the longest is the one that has the lowest Ohmic thermal losses. We note here that while magnetic field energy is converted into heat as  described by the term $j^2/\sigma$, this power is replenished by the fact that we enforce time-independent boundary conditions.

We note that the magnetospheric current closing through the crust is not related to the Hall current \citep{Goldreich:1992}. The latter is associated with the structure of the crustal magnetic field, and it can drive magnetic field evolution, especially for magnetic fields  above $10^{14}$~G. Here we assume for simplicity that the magnetic field in the crust is a pure dipole, and that the source of the field (i.e. the associated azimuthal electric current) lies interior to the crust.

\subsection{Boundary conditions}

We will integrate eq.~(\ref{MINIMdeq}) in a computational domain $r_{\rm in}\leq r \leq  r_{\rm out}$ and $0\leq \theta \leq  \theta_{\rm out}$ of the crust. We first need to specify $I(r,\theta)$ at the boundaries of the domain. We set the outer radial boundary at the radius of the star, namely $r_{\rm out}=r_{\rm ns}$, and the inner radial boundary at the inner crust radius $r_{\rm in}=0.9r_{\rm ns}$. Along the axis, $I(r, 0)=0$. $I(r_{\rm in}, \theta)=0$ as we assume that the current is contained within the crust, (this is a reasonable assumption due to the high resistivity that prevents the current from penetrating below the crust). $I(r,\theta>\theta_{\rm out})=0$. This is justified by the fact that the magnetspheric poloidal current flows only along open field lines, and no current flows in the dead zone. Inside the crust, the majority of the current stays below the polar cap region, and does not spread to much lower latitudes. Actually, as we shall see, part of the current spreads beyond the polar region, but as long as the latitudinal boundary $\theta_{\rm out}$ is taken to be sufficiently large, its particular value is not important. For computational convenience we take $\theta_{\rm out}=2\theta_{\rm pc}$, where  $\theta_{\rm pc}\approx (1.23\ r_{\rm ns}/r_{\rm lc})^{1/2}$ is the latitude of the footpoint of the last open field line on the star\footnote{We have also integrated eq.~(\ref{MINIMdeq}) with $\theta_{\rm out}=3\theta_{\rm pc}$ and the difference in the solution was minimal.}. Here, $r_{\rm lc}=c /\Omega$ is the radius of the light cylinder. The prefactor $1.23$ in the above expression is based on the most detailed numerical solution of the axisymmetric problem to date, namely that of \cite{Timokhin:2006}.

The distribution of electric current $I(r_{\rm out},\theta)$ along the surface of the neutron star is provided by the magnetospheric solution. This has been thoroughly investigated  by several authors \citep{Contopoulos:1999, Gruzinov:2005, Timokhin:2006}. Here we use the model with $x_0=0.992$ from  Fig.~3 of \cite{Timokhin:2006}. In that paper, $I$ is given as a function of $\Psi$, the poloidal magnetic flux, which on the surface of the neutron star is defined as $\Psi(\theta) =B r_{\rm ns}^2 \sin^2 \theta / 2 \pi$. Here and below, $B$ refers to the value of the magnetic field at the poles of the star. This allows us to obtain the distribution $I=I(r_{\rm out},\theta)$. Based on Timokhin's solution, the maximum value of $I$ is $I_{\rm max}=0.87 I_{\rm sm}$, where $I_{\rm sm}= 1.23 \times 2\pi B r_{\rm ns}^3 r_{\rm lc}^{-2}$ corresponds to a split-monopole solution with the same amount of open magnetic flux \citep{Michel:1973}. Therefore, the maximum value used in the present work is $I_{\rm max}=1.07 \times 2\pi B r_{\rm ns}^3 r_{\rm lc}^{-2}$. We implemented that by sampling the corresponding curve ανδ constructing a table with $\Psi~-~I$ pairs. \cite{Gralla:2016} have shown that this solution is approximated to high accuracy by the polynomial expression
\begin{eqnarray}
I(\Psi) = \frac{\Psi}{\Psi_0}\left[2-\frac{\Psi}{\Psi_0}-\frac{1}{5}\left(\frac{\Psi}{\Psi_0}\right)^3\right],
\end{eqnarray}
where $\Psi_0=1.23$ to agree with the normalisation adopted above. In our approach we have solved the equation using expressions for the boundary condition, noting a deviation between the solutions of $\sim 1\%$.

Outside the polar cap, $I(r_{\rm out}, \theta)$ drops to zero as a step function. We have smoothened this abrupt drop within a layer of width $0.01 \theta_{\rm pc}$. This smooths out the current density  near the surface of the star, but leaves the flow of current deeper in the crust mostly unaffected.

\subsection{Neutron star parameters}

\begin{figure}
  \includegraphics[width=0.5\textwidth]{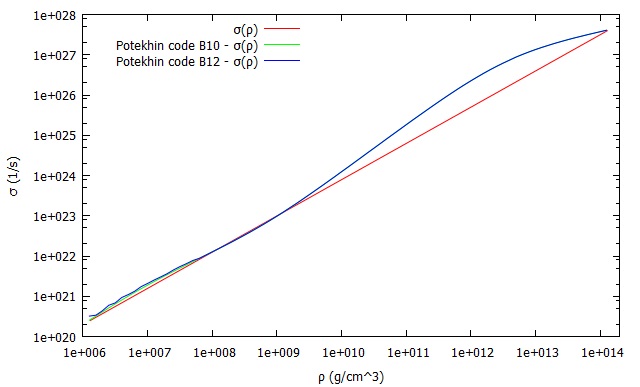}
     \caption{Conductivity function $\sigma(\rho)$ (eq.~\ref{condeq}) in comparison to respective results from the Potekhin code \protect\cite{Potekhin:2015}.}
      \label{condfig}
\end{figure}

 We consider a neutron star radius $r_{\rm ns}=10$~km and we adopt a typical ground state structure for the crust as described in \cite{Chamel:2008} (\S~3 see Figure~4).
Atoms are fully ionized at mass densities higher than about $\rho \sim 10^4$~g~cm$^{-3}$. The so-called ocean extends up to densities of $10^{6}$~g~cm$^{-3}$. This forms a layer from about a few meters up to $100$~m, depending on the temperature of the neutron star \citep{Potekhin:2015}. Below the ocean, the outer crust consists of a body-centered iron $^{56}$Fe cubic lattice with the composition of the nuclei becoming more neutron rich as a result of electron capture. The inner crust region extends from $\rho_{\rm nd} \sim 4 \times 10^{11}$ to about $10^{14}$~g~cm$^{-3}$. At the bottom of the crust, some calculations predict various ``pasta'' phases of non-spherical nuclei, such as slabs or cylinders  \citep{Horowitz:2015}. Such pasta layer are believed to be highly resistive due to the anisotropic structure of the lattice and a low electron fraction \citep{Pons:2013}. In the present work, we consider only the part of the crust ranging from the base of the ocean down to the crust-core boundary. This corresponds to densities ranging from $\rho_{\rm out}=1.3 \times 10^{6}$~g~cm$^{-3}$ to $\rho_{\rm in}=1.3\times 10^{14}$~g~cm$^{-3}$ at the outer and inner crust boundary, respectively. Finally, we express the density of the crust as a function of the depth from the neutron star surface \citep{Chamel:2008} with the following analytical expression 
\begin{eqnarray}
\rho= \left( 1 + \left(\frac{r_{\rm ns}-r}{r_{\rm ns}-r_{\rm in}}\right)^4\frac{\rho_{\rm in}}{\rho_{\rm out}} \right) \rho_{\rm out} \, .
\label{denfun}
\end{eqnarray}

The expression for the electric conductivity $\sigma(\rho)$ is taken from analytical fits of the numerical solutions obtained using the codes developed by \cite{Potekhin:2015}\footnote{The codes are available at {\it http://www.ioffe.ru/astro/conduct/index.html}}.
We have chosen the following set of parameters: ion charge  number (atomic number) Z$=26$, mass number A$=56$, 
impurity parameter Z$_{\rm imp}=0.1$, range of densities $10^{6}\leq \rho\leq 10^{14}$~g~cm$^{-3}$, and temperature $T=10^{7}$~K. We have experimented with  two magnetic field values, $B=10^{10}$~G and $10^{12}$~G.
The differences between the two conductivity estimates are minimal (see the green and blue curves in Fig.~\ref{condfig}), and we approximate them by the following analytical power-law expression
\begin{eqnarray}
\sigma(r)=\sigma_{\rm out} \, \left(\frac{\rho(r)}{\rho(r_{\rm ns})}\right)^{9/10} \, ,
\label{condeq}
\end{eqnarray}
with $\sigma_{\rm out}=2.5\times 10^{20}$~s$^{-1}$ (red line in Fig.~\ref{condfig}). We note the small deviation between the expressions derived in \cite{Potekhin:2015} and the analytical expression used here. We have verified that they have minimal impact on the electric current flow inside the crust (less than $1\%$ deviation). To assess the importance of the conductivity profile, we also integrated eq.~(\ref{MINIMdeq}) for a constant conductivity ($\sigma=10^{24}$ s$^{-1}$) and we report the differences below.

\section{Results}

We solve eq.~(\ref{MINIMdeq}) with the Gauss-Seidel numerical method. The algorithm for this elliptic solver is provided in Numerical Recipes \citep{Press:1988}. We initialize the scheme with a trial distribution $I(r, \theta)$ and we repeat the iterative procedure until convergence is achieved. We implemented an $r-\theta$ numerical grid with a uniform resolution of $160 \times 400$, and we find that the solution converges after $10^6$ iterations.

Based on the solutions that we have derived, we can evaluate the torque exerted on the pulsar, the corresponding stresses, and the Ohmic heating. In order to calculate these quantities, we assume that the internal magnetic field is a dipole, namely 
\begin{eqnarray}
{\bf B}_p(r, \theta) = B \, r^3_{\rm ns} \left( \frac{\cos \theta}{r^3} {\bf {\hat{r}}} + \frac{\sin \theta}{2r^3} {\bf{\bm{\hat{ \theta}}}} \right)\,. 
\label{Beq}
\end{eqnarray}
First, we calculate the Lorentz force per unit volume using eq.~(\ref{CURRENT}), 
\begin{eqnarray}
{\bf F}_L(r, \theta) &=&~\frac{1}{c}~ {\bf j} \times {\bf B}_p \nonumber \\
&=& \frac{B r_{\rm ns}^3}{4\pi r^4} \left( \frac{1}{2r} \frac{\partial I}{\partial \theta} + \frac{\cos \theta}{\sin \theta} \frac{\partial I}{\partial r} \right) {\bf{\bm{\hat{\phi}}}}\, .
\label{FLeq}
\end{eqnarray}
Here we have to note that because of $ {\bf E}= \left( r_{ns}/r_{lc} \right){\bf B}_p << {\bf B}_p$, the electrostatic term $\rho_e {\bf E}$, which exists  in eq.~(\ref{FLeq}) is $\left( r_{ns}/r_{lc} \right)^2$ times smaller than the calculated one, so in limit of our numerical error is negligible.

Then, we calculate the torque per unit volume
\begin{eqnarray}
{\bf N}(r, \theta) &=& {\bf r} \times {\bf F}_L \nonumber \\
&=& \frac{B r_{\rm ns}^3}{4\pi r^3}\left(- \frac{1}{2r} \frac{\partial I}{\partial \theta} - \frac{\cos \theta}{\sin \theta} \frac{\partial I}{\partial r} \right) {\bf{\bm{\hat{\theta}}}}\,.
\label{toreq}
\end{eqnarray}
Because of axial symmetry only the torque component parallel to the axis of symmetry is non-zero. Thus, the total torque is given by the integral 
\begin{eqnarray}
N_{\rm tot}=\int_V (N_r\cos\theta-N_\theta\sin\theta) \, dV \, .
\label{itoreq}
\end{eqnarray}
Finally, we calculate the components of the Maxwell stresses. The diagonal components correspond to pressure terms, while the off-diagonal components lead to the deformation of the crust due to shear-stresses. $M_{r\theta}$ involves the $B_r$ and $B_\theta$ components which are not due to the pulsar spin-down current. 
\begin{eqnarray}
M_{r\phi} = \frac{B_r(r,\theta) \, B_{\phi}(r,\theta)}{4 \pi}\ \qquad \mbox{and}\ \qquad 
M_{\theta \phi} = \frac{B_\theta(r,\theta) \, B_{\phi}(r,\theta)}{4 \pi}
\label{eqstr}
\end{eqnarray}
are due to the spin-down current which is associated with a toroidal magnetic field $B_{\phi}$ (see eq.~\ref{Bfeq}). 
The breaking stress limit of the crust is 
\begin{eqnarray}
\tau_{br} = \left( 0.0195 - \frac{1.27}{\Gamma - 71} \right) n_i \frac{Z^2 e^2}{\alpha}
\label{eqellim}
\end{eqnarray}
\citep{Chugunov:2010}, where $\Gamma = Z^2 e^2 /\left( \alpha k_B T \right)$ is the Coulomb coupling parameter, $\alpha =\left[ 3/ \left(4 \pi n_\mathrm{i} \right) \right]^{1/3}$ is the ion sphere radius, $n_\mathrm{i}$ is the ion number density, $k_\mathrm{B}$ is the Boltzmann constant and $e$ is the electron charge.  A Maxwell stress comparable to $\tau_{br}$ may lead to crust yielding and deformation. 
We evaluate the breaking stress for densities in the range $\rho_{\rm nd}< \rho\leq \rho_{\rm in}$ using the table from \cite{Douchin:2001} and for densities $\rho_{\rm out} \leq \rho \leq \rho_{\rm nd}$ using the results of \cite{Haensel:1994}. The breaking stress at the base of the crust is $\tau_{br}(\rho =10^{14}$g cm$^{-3})= 2\times 10^{29}$~dyn~cm$^{-2}$, at the neutron drip point $\tau_{br}(\rho=4\times 10^{11}$g cm$^{-3})=1.4\times 10^{27}$~dyn cm$^{-2}$ and at the base of the ocean $\tau_{br}(\rho =10^{6}$~g cm$^{-3})=10^{20}$~dyn cm$^{-2}$. These results at the base of the crust and at the neutron drip point are consistent with the estimates of \cite{Cumming:2004, Lander:2019}. The value near the upper boundary of our integration domain (the base of the ocean) depends on temperature through the Coulomb coupling parameter $\Gamma$.

The most important parameter of the problem is the pulsar period. This determines the size of the polar cap, and thus the boundary conditions on the surface. We have integrated eq.~(\ref{MINIMdeq}) for 5 different choices of the period $P=0.01,~0.1,~1,~5,~7.5$ s, thus exploring configurations that range from a rapidly rotating millisecond pulsar to a slowly rotating magnetar. The flow pattern of the electric current in the crust is independent of the strength of the magnetic field. Nevertheless, the physical quantities that we evaluate below depend on it. For this reason, we have assigned realistic values to the magnetic field ranging from $10^{10}$ to $10^{15}$ G to allow a direct comparison. The combinations employed are shown in detail in Table~\ref{tres}. For each magnetic model we integrate eq.~(\ref{MINIMdeq}) both for a constant and a variable conductivity given by eq.~(\ref{condeq}). We also report the deepest point reached by the electric current flow line that corresponds to $I(r,\theta)=I_{\rm max}/2$. This yields an estimate of the electric current penetration 'half-depth', namely how deep $50\%$ of the current reaches inside the star. We evaluate the Ohmic power using eq.~(\ref{OHMIC_POWER}). 

As a consistency test, we calculate the torque exerted on the star by integrating eq.~(\ref{itoreq}), and comparing it with the spin-down torque of an aligned rotator in the force-free approximation appropriately corrected \citep{SpitkovskyContopoulos:2006}
\begin{eqnarray}
N_{\rm align} = 0.94\times \frac{2}{3} \frac{\Omega}{c} \left(\frac{1.23 r_{\rm ns}^3 B}{r_{\rm lc}} \right)^2 \, .
\label{naleq}
\end{eqnarray}
The correction factor of $0.94$ is due to the integral of the electromagnetic luminosity \citep{Gralla:2016}. Once this is taken into account the volume integral of the spin-down torque $N_{tot}$ and $N_{\rm align}$ expression are for most models within a $2\%$ difference from each other. The results obtained using the boundary condition from the numerical solution of \cite{Timokhin:2006} and the polynomial fit of \cite{Gralla:2016} give the same results within $2\%$ of each other, as expected since the two solutions agree at this level of accuracy.

\begin{table*}
\begin{center}
\caption{Summary of the models studied. The first column is the name of the model. $P$ and $\dot{P}$ are the period and period derivative, $B$ is the spin-down dipole magnetic field of an orthogonal rotator in vacuum corresponding to the period and period derivative mentioned before, $\theta_{\rm pc}$ is the semi-opening angle of the polar cap, the depth is the lowest value of $r$ for $I=\frac{1}{2}I_{\rm max}$, $N_{\rm tot,~T}$ is the torque obtained through the integration of equation \ref{toreq} using the boundary condition from \protect\cite{Timokhin:2006} and $N_{\rm tot,~G}$ is the torque using the polynomial approximation of \protect\cite{Gralla:2016}, $N_{\rm align}$ is the torque evaluated through equation \ref{naleq}, $P_{\rm Ohm}$ is the total Ohmic losses using the polynomial approximation of \protect\cite{Gralla:2016}. The last column indicates whether the conductivity is set equal to a constant (c) or depends on radius according to equation \ref{condeq}  (v).} 
\begin{tabular}{ |c|c|c|c|c|c|c|c|c|c|c| }
\hline
Model & $P$ & $\dot{P}$ & $B$ & $\theta_{\rm pc}$ & depth & $N_{\rm tot, ~T}$&$N_{\rm tot, ~G}$ & $N_{\rm align}$ & $P_{\rm Ohm}$ &$\sigma$ \\
& (s) & & (G) & (deg)& ($\times 100$ cm) & (dyn~cm) & (dyn~cm)& (dyn~cm) & (erg/s) & \\
 \hline
 A1 & 0.01 & 9.77e-18 & $10^{10}$ &  9.24 & 740 & 8.7e32& 8.5e32 & 8.7e32  & 1.2e19 & v \\ 
 A2 & 0.01 & 9.77e-18 & $10^{10}$ & 9.24 & 280 & 8.8e32& 8.6e32 &  8.7e32 & 2.5e16 & c \\
\hline
 B1 & 0.1 & 9.77e-15 & $10^{12}$ & 2.91 & 310 & 8.8e32& 8.6e33 & 8.7e33 & 4.9e19 & v \\ 
 B2 & 0.1 & 9.77e-15 & $10^{12}$ & 2.91 & 100 & 8.8e32& 8.6e33 & 8.7e33 & 6.6e16 & c \\
\hline
 C1 & 1 & 9.77e-16 & $10^{12}$ & 0.92 & 100 & 8.8e32 & 8.7e30 & 8.7e30 & 2.1e16 & v \\ 
 C2 & 1 & 9.77e-16 & $10^{12}$ & 0.92 & 40& 8.7e32 & 8.5e30 & 8.7e30 & 1.7e13 & c \\
\hline
 D1 & 5 & 1.95e-12 & $10^{14}$ & 0.41 & 50 & 7.1e32 & 7.1e32 & 7.0e32 & 9.4e17 & v \\ 
 D2 & 5 & 1.95e-12 & $10^{14}$ & 0.41 & 20 & 6.8e32& 6.6e32 & 7.0e32 & 4.7e14 & c \\
\hline
SGR 1806$-$20 & 7.5 & 4.95e-12 & $2\times 10^{15}$ & 0.33 & 40 & 8.4e32 & 8.0e34 & 8.2e34 & 6.2e19 & v \\
\hline
\end{tabular}
\label{tres}
\end{center}
\end{table*}

\section {Discussion}

\subsection{Electric current flowlines and Joule heating}

\begin{figure*}
 A1 \includegraphics[width=0.45\textwidth]{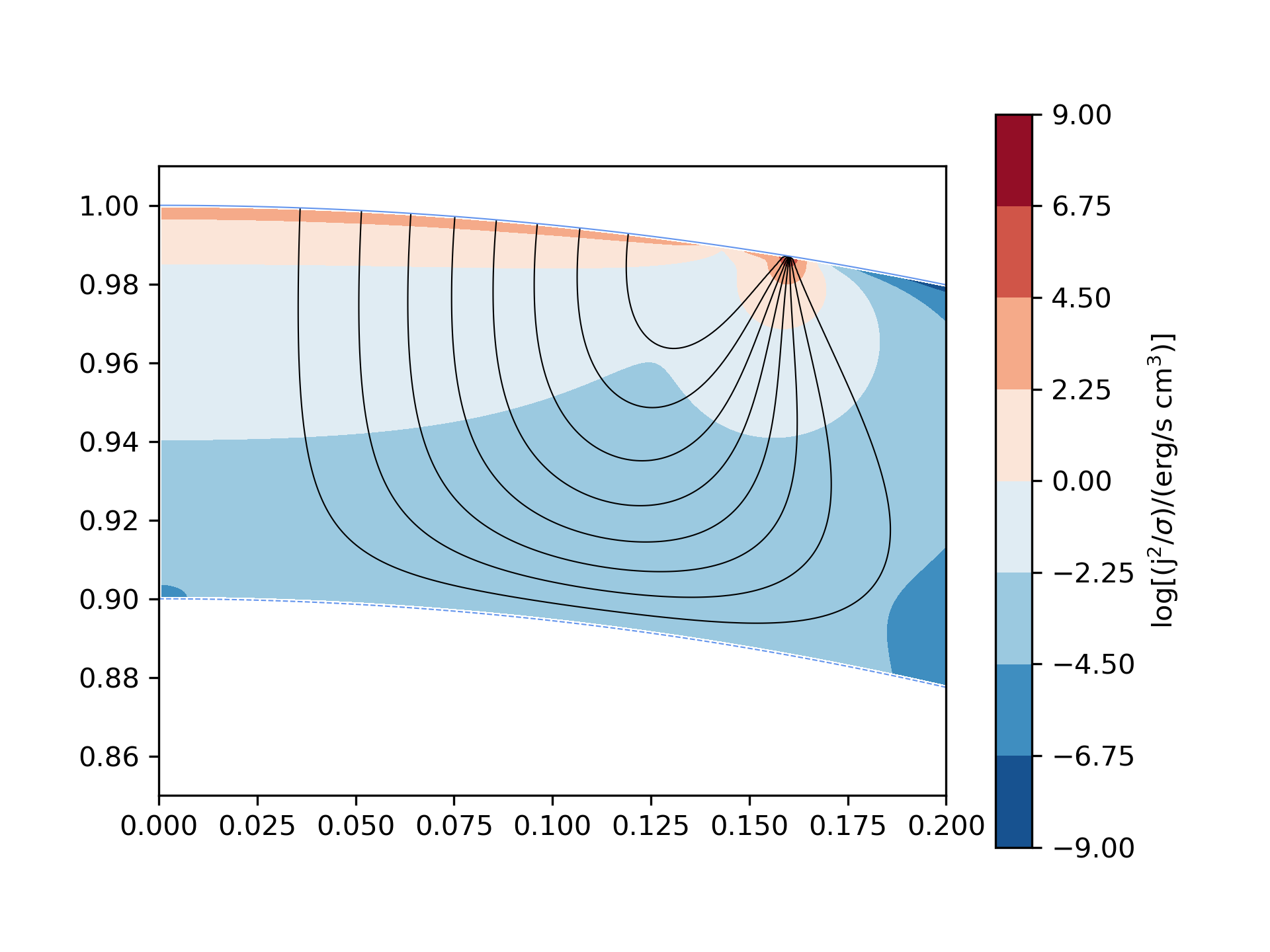}
 B1   \includegraphics[width=0.45\textwidth]{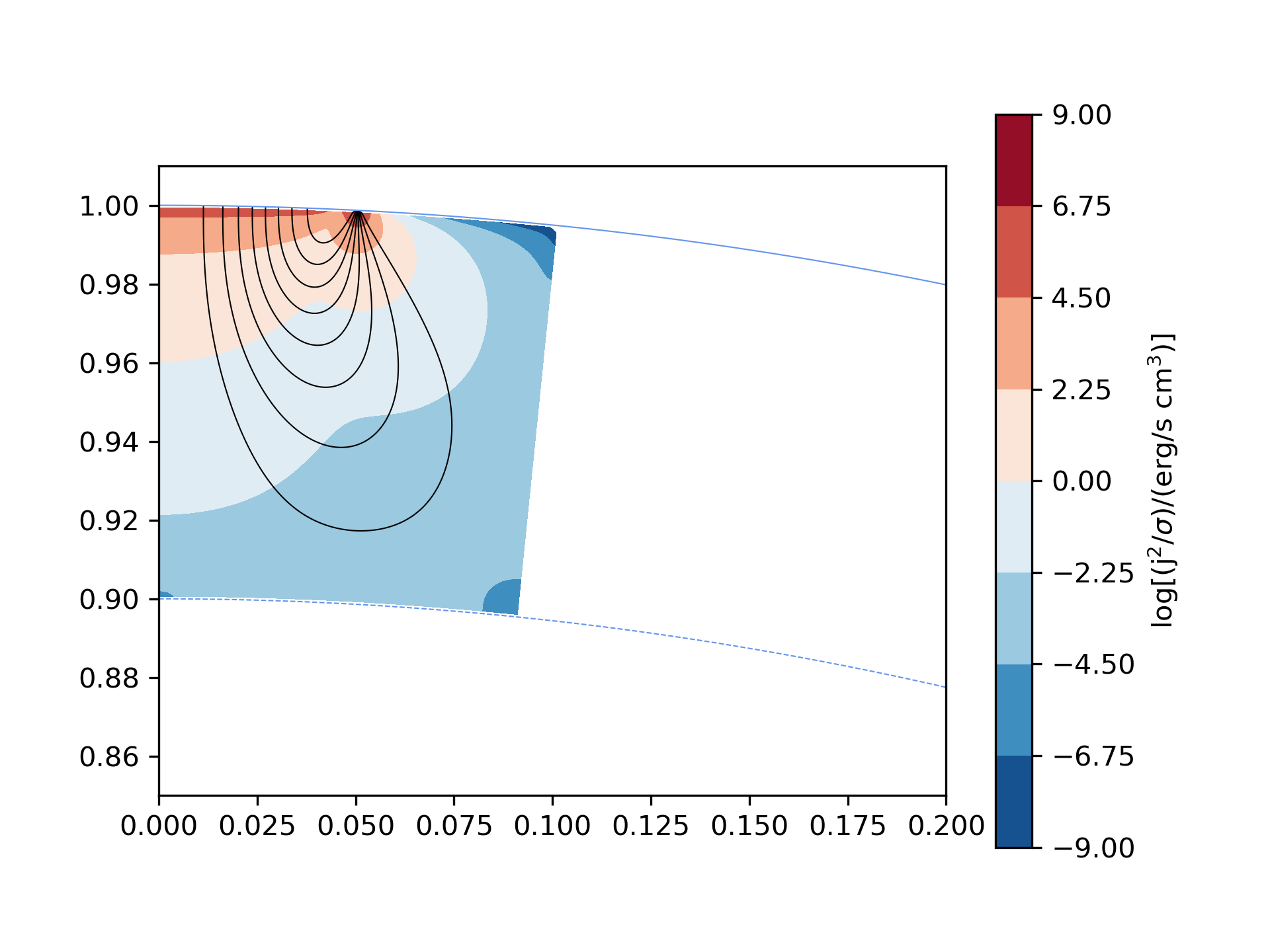}\\
 C1     \includegraphics[width=0.45\textwidth]{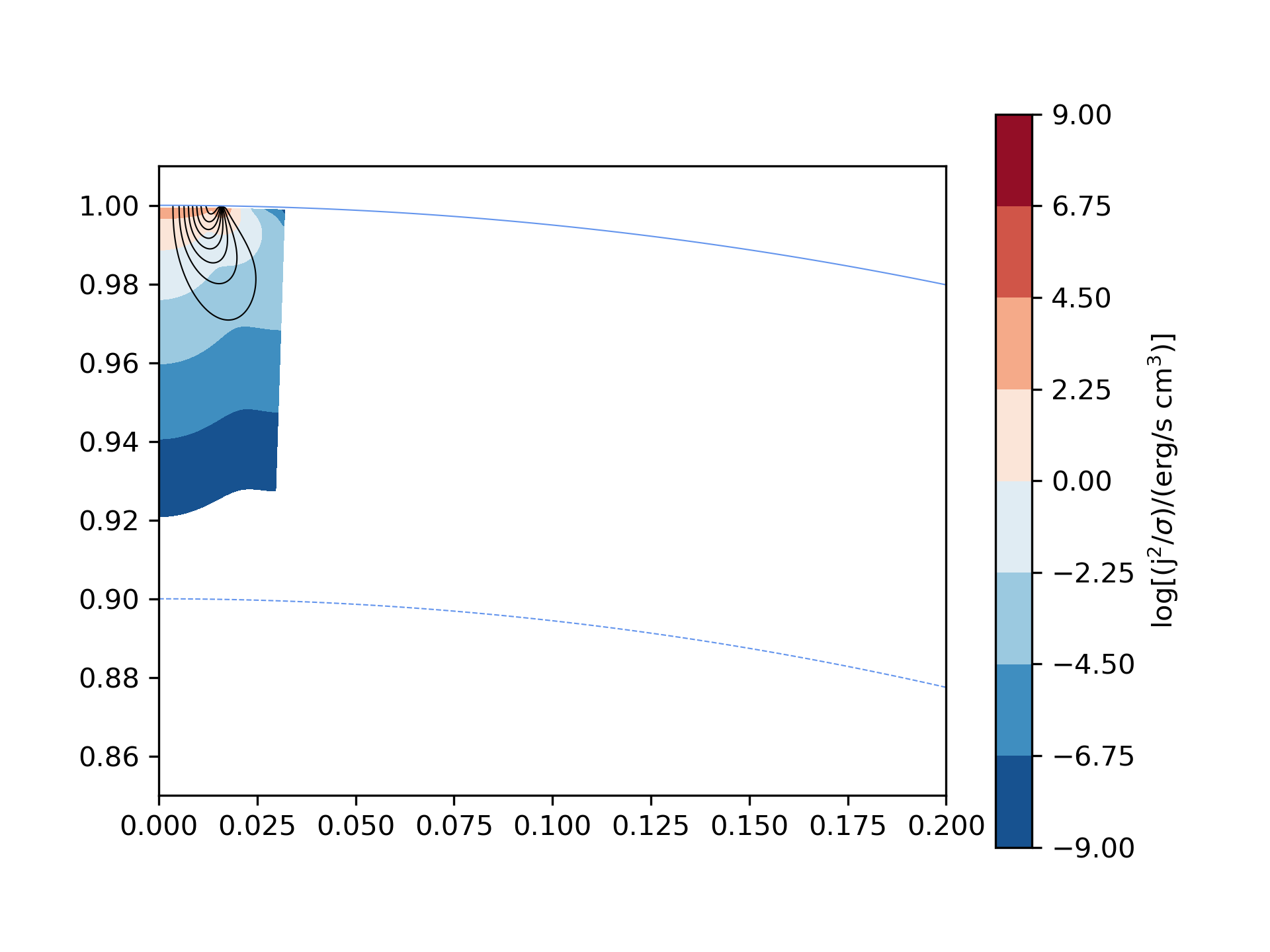}
 D1 \includegraphics[width=0.45\textwidth]{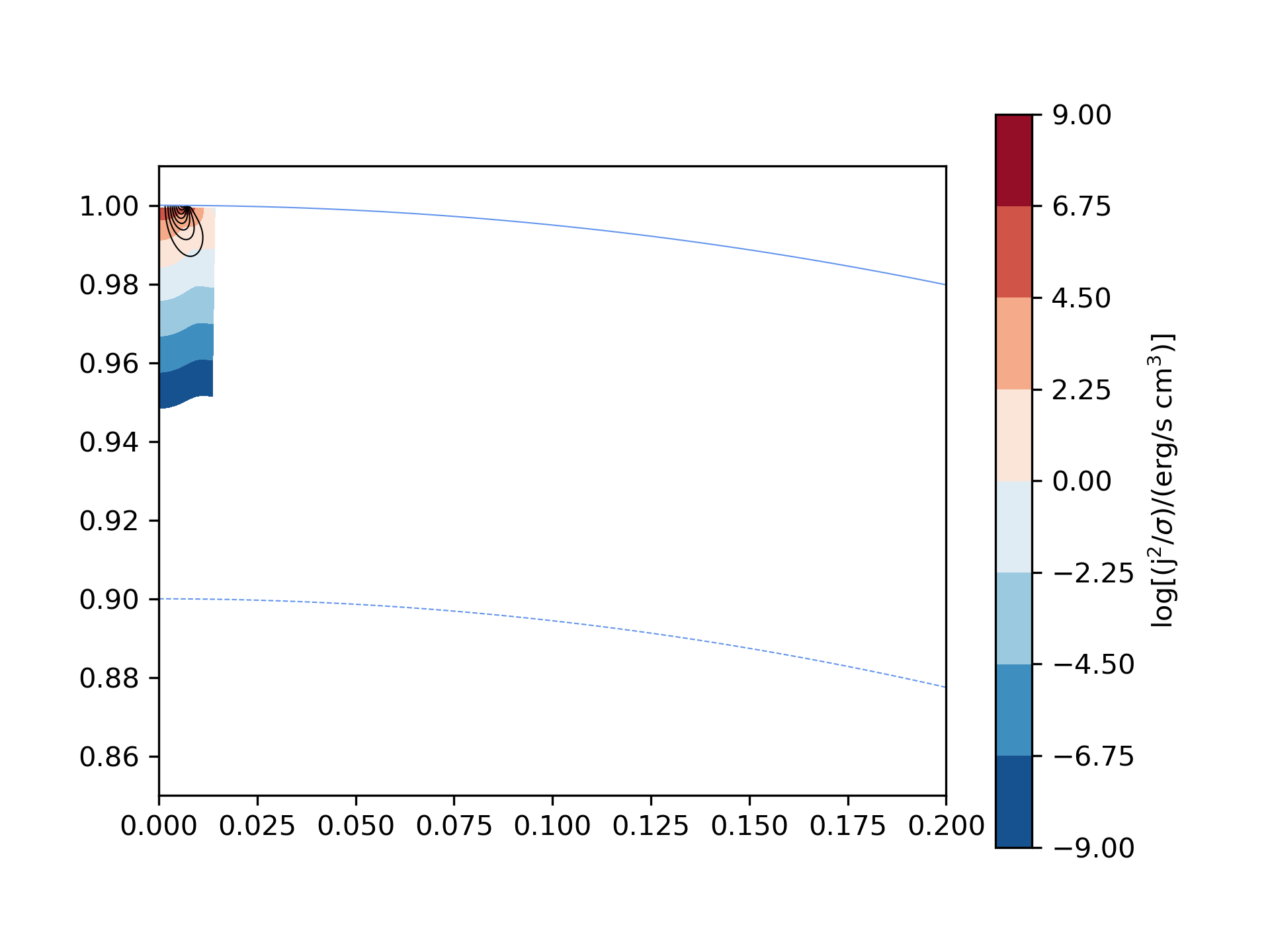}
    \caption{Plots of electric current flow-lines in black and Joule heating per unit volume in color for models A1, B1, C1, D1. Horizontal and vertical distances in units of $r_{\rm ns}$. Continuous thin line: outer stellar surface. Dotted thin line: base of the crust.}
     \label{Fig:Ohm}
\end{figure*}

\begin{figure}
  \includegraphics[width=0.45\textwidth]{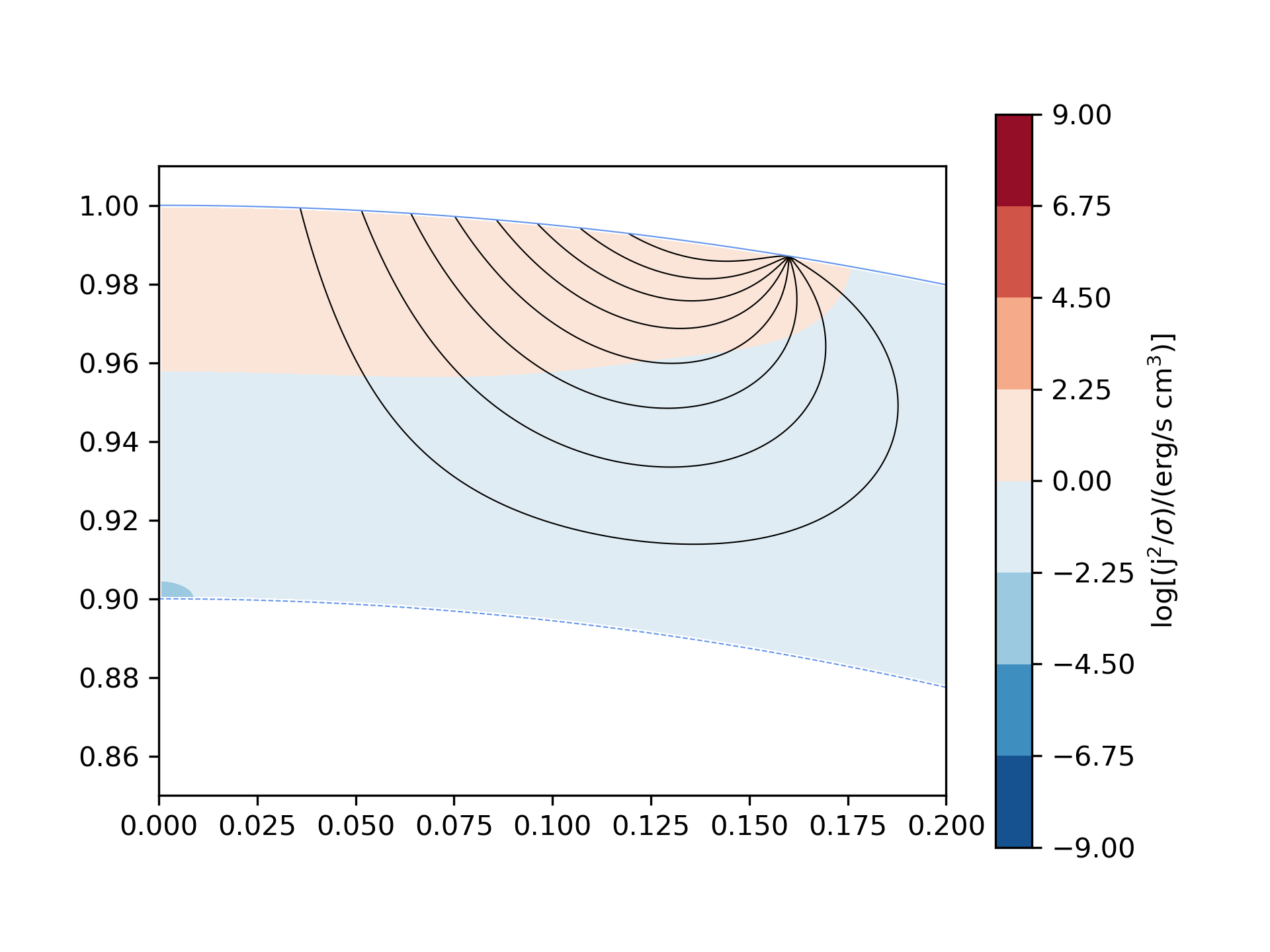}
    \caption{Electric current flow-lines in black and Joule heating per unit volume in color for models A2. Here the conductivity is constant and the current remains at a shallower depth compared to model A1 (Fig.~\ref{Fig:Ohm} top left panel).}
 \label{Fig:Ohm_c}
 \end{figure}

In all models, there is a significant penetration of the electric current in the crust (Fig.~\ref{Fig:Ohm}). This is rather prominent in model A (millisecond pulsar) where the electric current practically reaches the base of the crust, with $50\%$ of the current reaching depths greater than $740$~m. Here the solution is affected by the boundary condition enforced at the base of the crust that does not allow the electric current to proceed any deeper. If this constraint were to be relaxed assuming the rest of the star had a similar conductivity, the current would formally reach into the core. In models B, C, D and SGR 1806$-$20, the current travels to a much smaller depth which scales with the radius of the polar cap. In these models the polar cap radii are smaller than the crust radius and the boundary condition at the base of the crust does not play any significant role. We remark further that a constant conductivity calculation yields a depth attained by the current approximately equal to $0.4$ times that of a realistic conductivity calculation (Fig.~\ref{Fig:Ohm_c}). 

The paths of the electric current illustrate how the minimisation of Ohmic losses is achieved. Ohmic losses are larger for higher electric current densities, yet for lower ones, the same total current imposed on the boundary has to travel a longer distance inside the crust that eventually leads to a larger integration volume. Thus, if the conductivity is kept constant, the current will follow a path compromising these two effects. Once the conductivity varies with depth, the current will travel even deeper as this will allow it to cross a region of lower resistivity and thus suffer less Ohmic losses, despite the total path being longer. Joule heating is higher near the surface and decreases towards the base of the crust. This variation is more pronounced for the realistic conductivity profile, as there the conductivity increases by several orders of magnitude as one approaches the base of the crust. The maximum Joule heating occurs at the rim of the polar cap. This is because the bulk of the current enters the crust through this location leading to formally infinite current density. The total Joule heating scales with the magnetic field as $\propto B^2$, and with period as $\propto P^{-2.5}$.

We note that the total Ohmic losses occuring in the crust are negligible compared to the total radiated spin-down power (typically 10 orders of magnitude smaller). This implies that the coupling between the crust and the electric current is strong. Furthermore, while the coupling with the crust is essential for the pulsar spin-down, its effect on the global pulsar electric circuit is minimal.

\subsection{Torque}

\begin{figure*}
 A1 \includegraphics[width=0.45\textwidth]{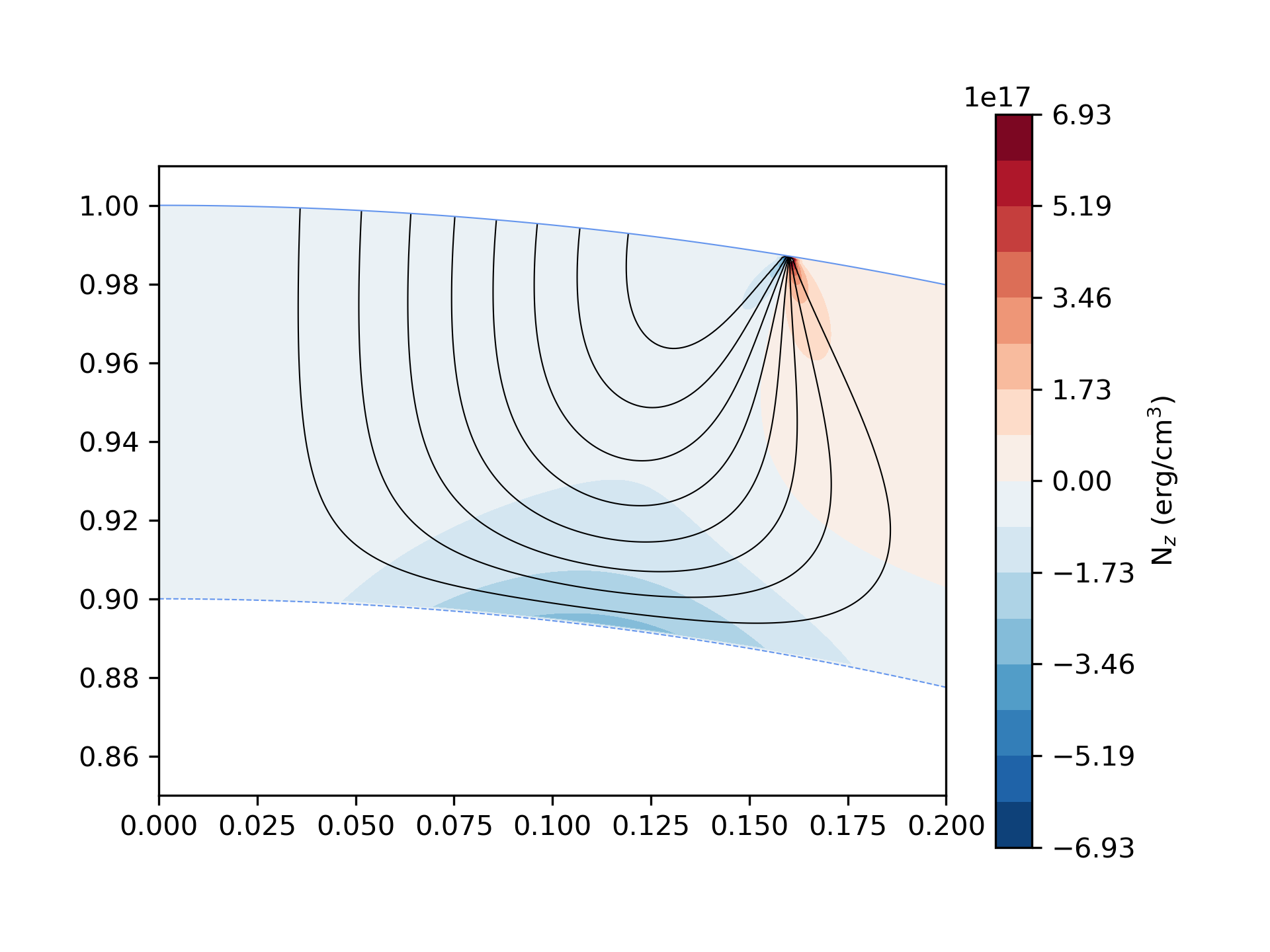}
 B1   \includegraphics[width=0.45\textwidth]{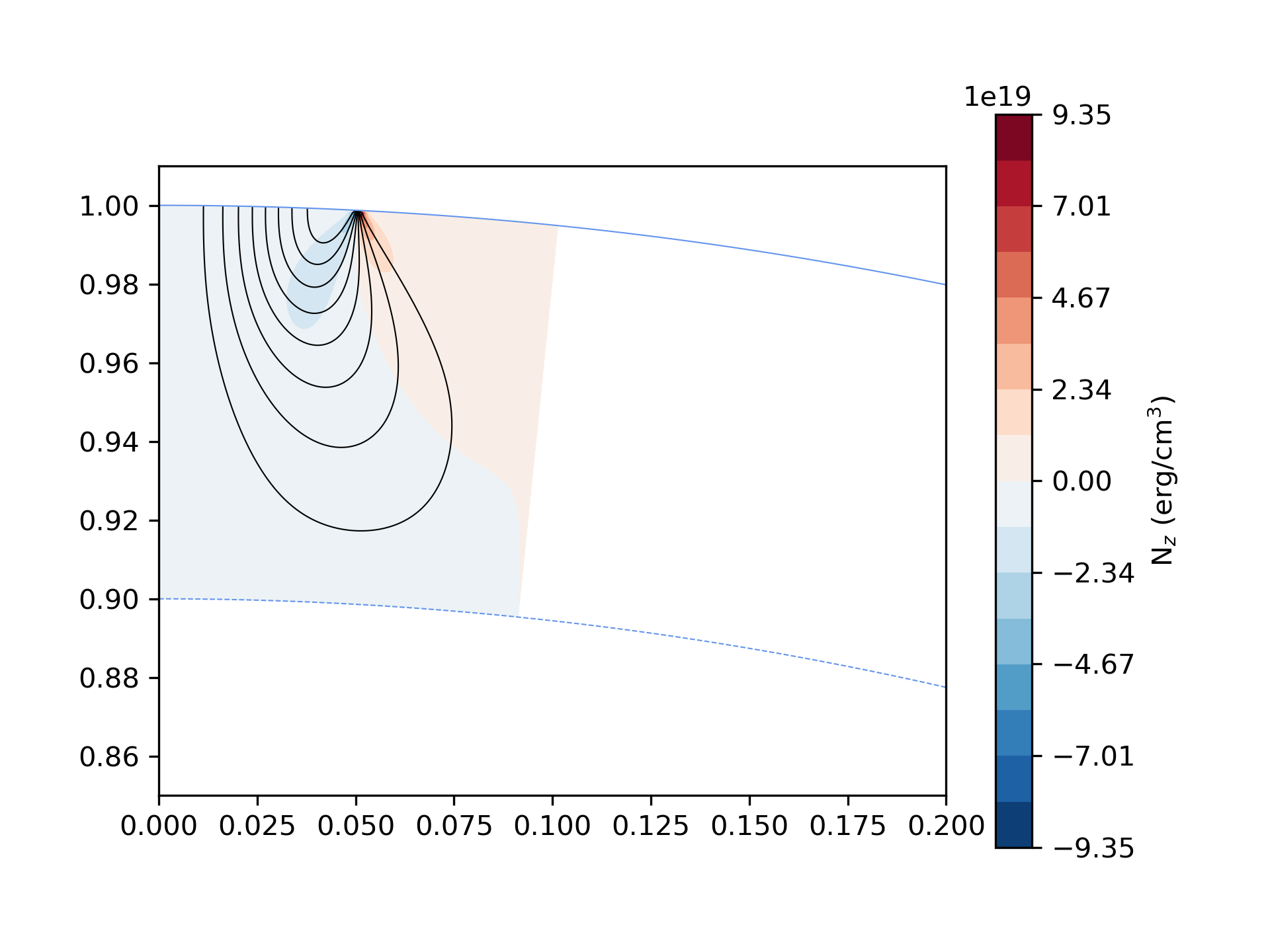}\\
 C1     \includegraphics[width=0.45\textwidth]{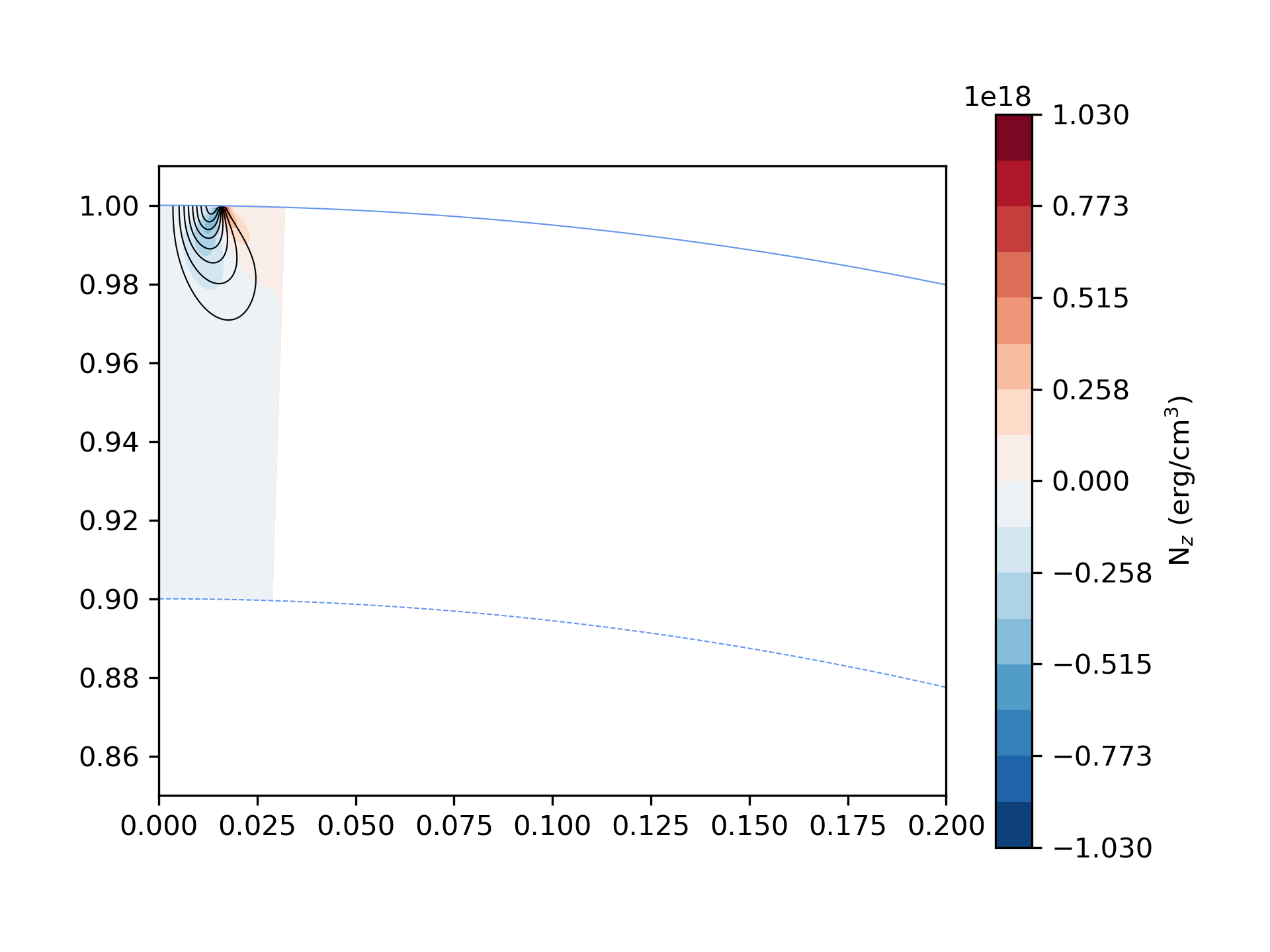}
 D1 \includegraphics[width=0.45\textwidth]{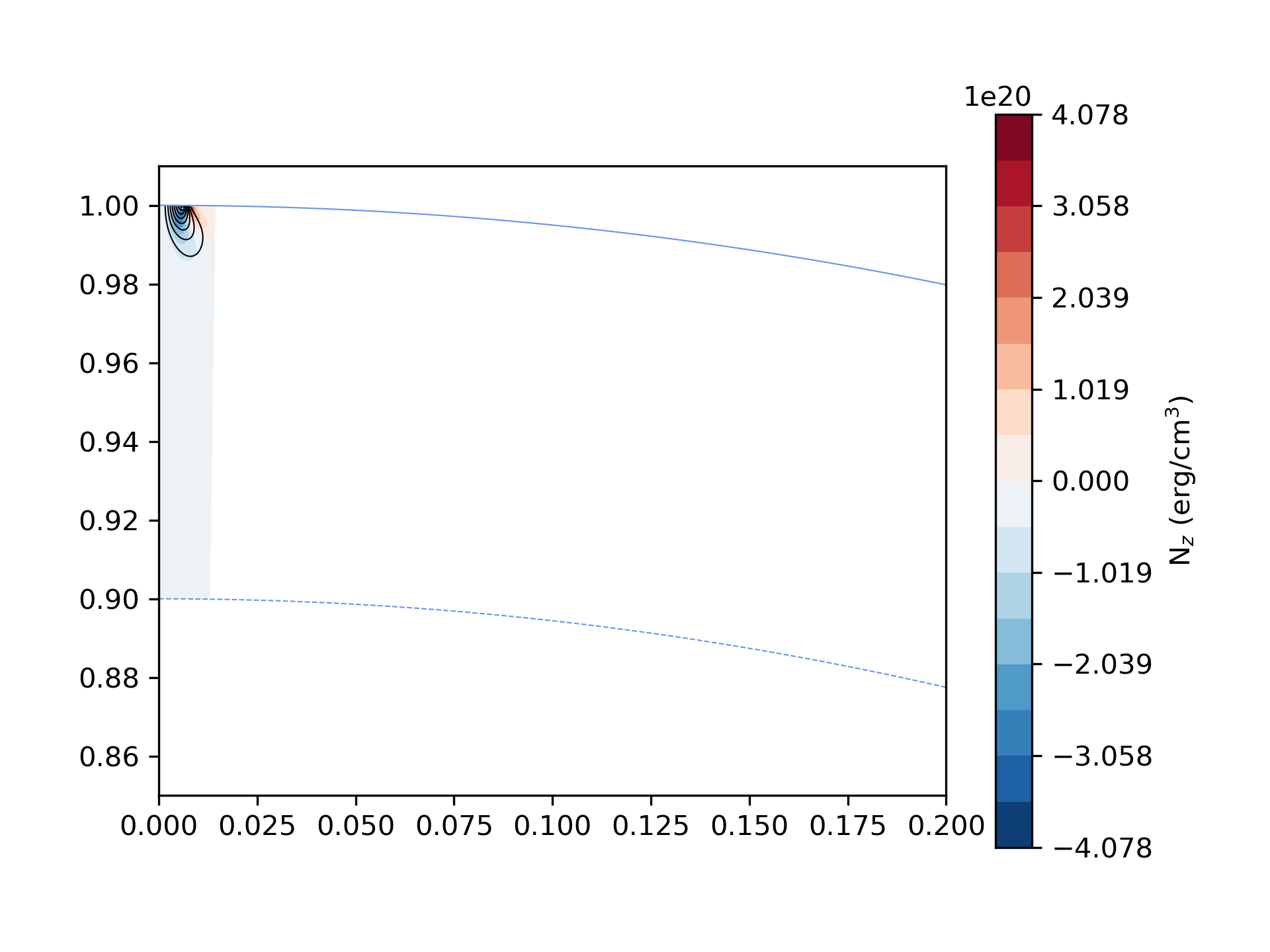}
    \caption{Plots of electric current flow-lines in black and torque per unit volume in color for models A1, B1, C1, D1. }
 \label{Fig:Torque}
\end{figure*}

As we saw in the previous section, the global torque approximates quite accurately the spin-down torque calculated through the magnetosphere. An interesting point here, is that there is a significant amount of localized spin-up torque that is of course overwhelmed by the spin-down torque. The reversal occurs along the surface $\partial I/\partial r =0$ where the current flow-lines become radial. The spin-up torque is mostly exerted on the part of the star where $\theta>\theta_{\rm pc}$, due to the spreading of the electric current at latitudes smaller than that of the polar cap  (see Fig.~\ref{Fig:Torque}).  

The maximum torque per unit volume occurs in the region below the rim of the polar cap, and is due to the high electric current density there. In the rapidly spinning model (A1), the bulk of the spin-down torque is exerted close to the base of the crust. For slower rotating models, the bulk of the torque is exerted closer to the surface.

\subsection{Maxwell Stresses}

\begin{figure*}
 A1 \includegraphics[width=0.45\textwidth]{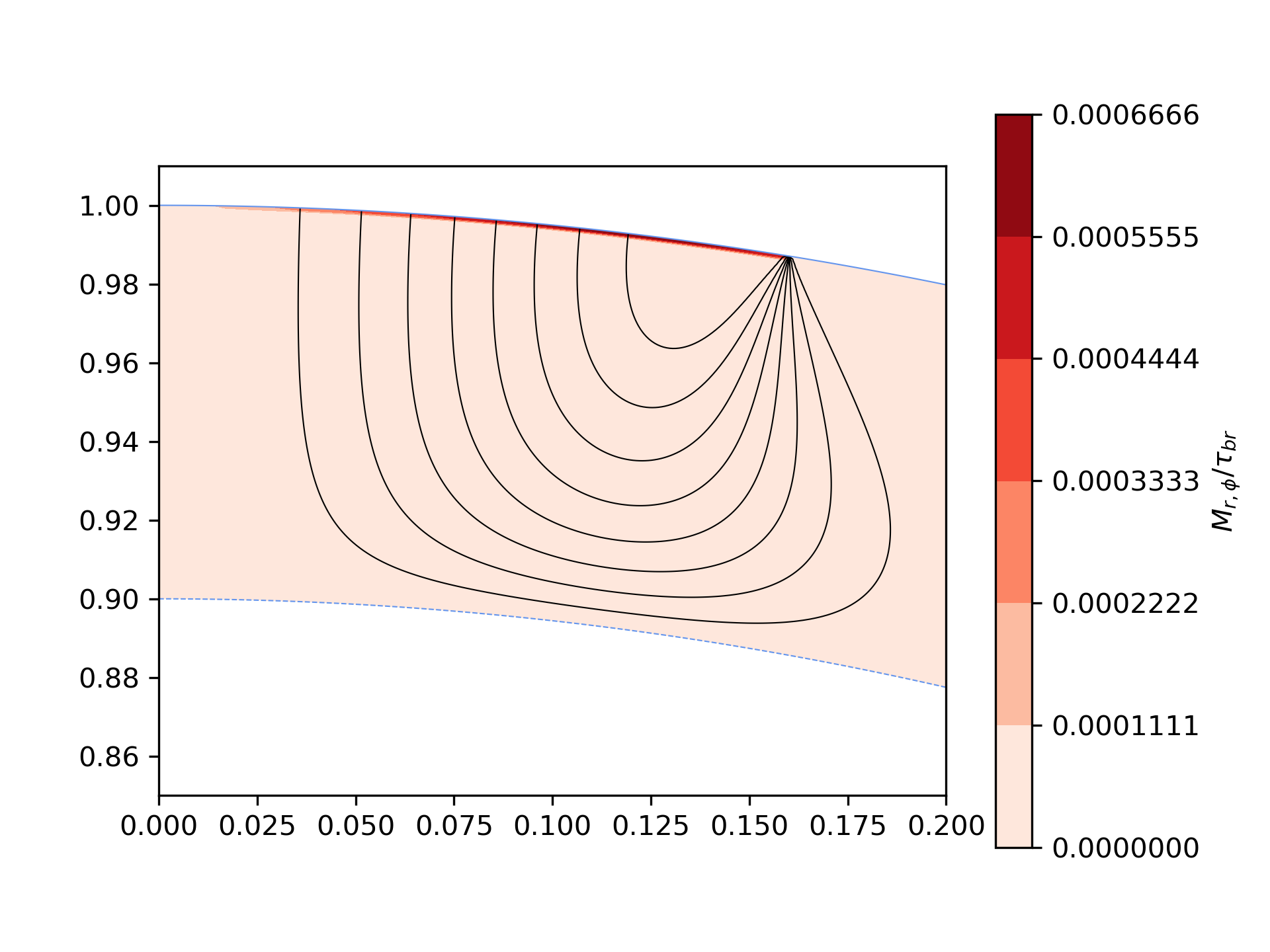}
 B1   \includegraphics[width=0.45\textwidth]{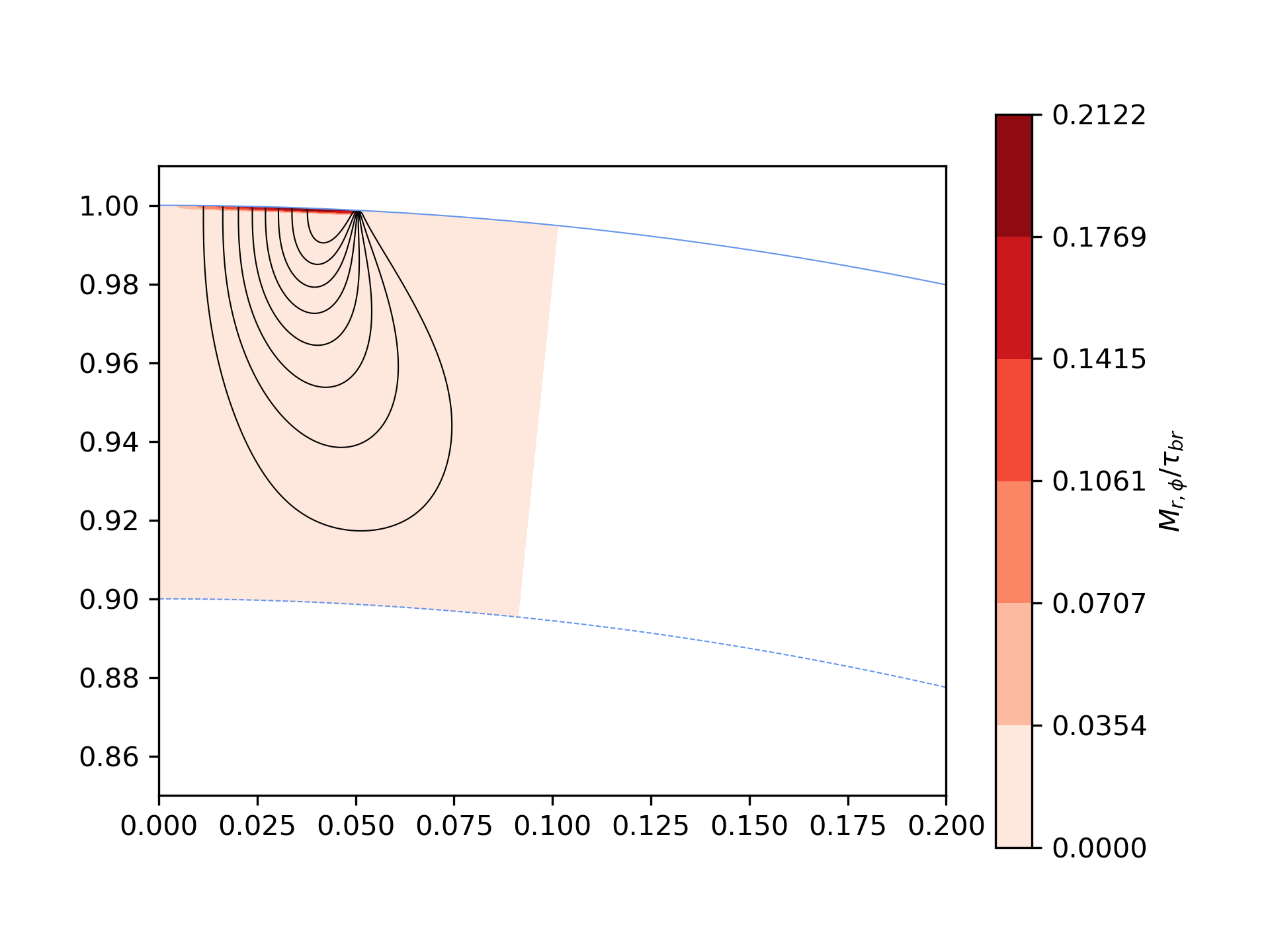}\\
 C1     \includegraphics[width=0.45\textwidth]{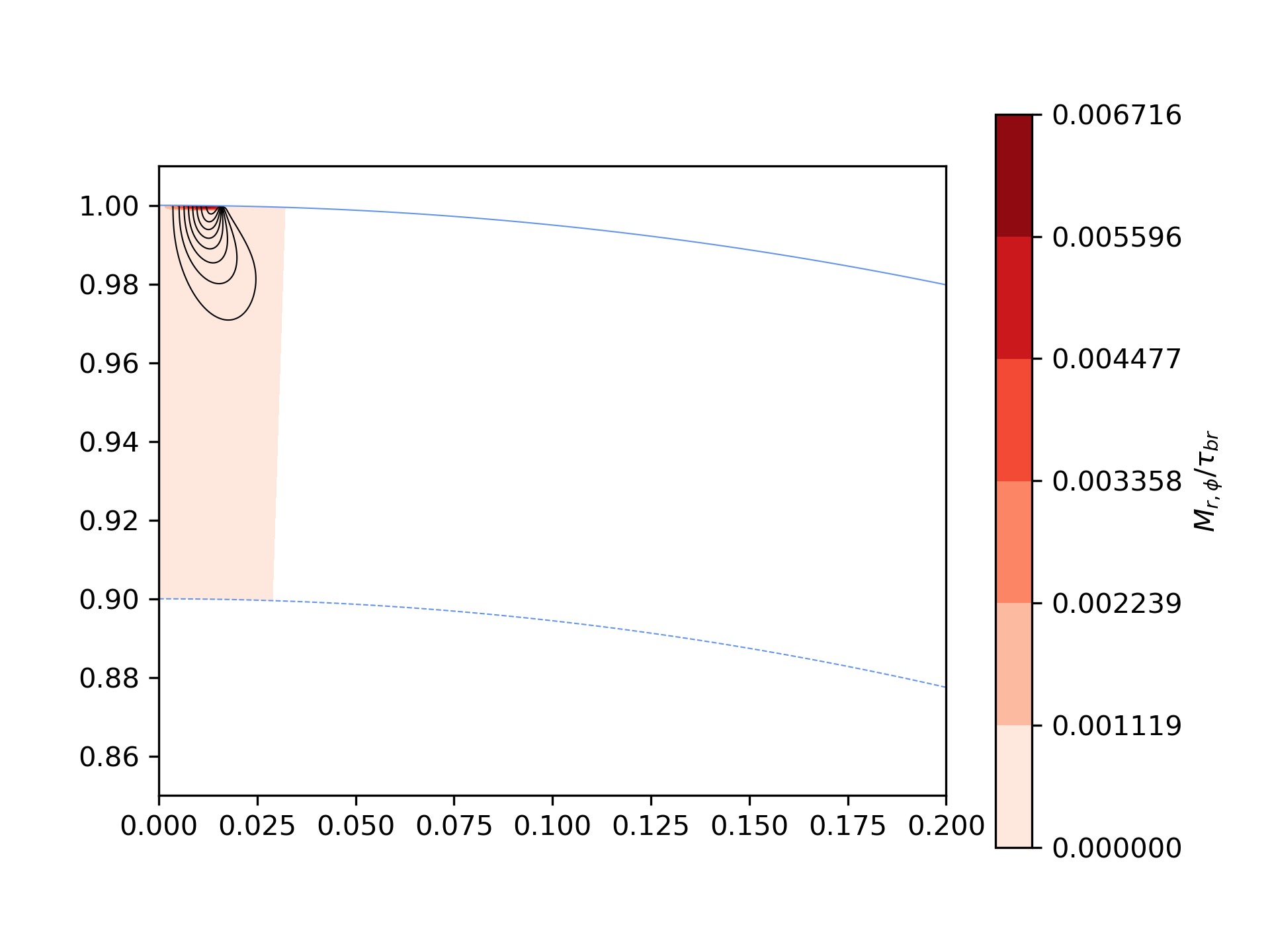}
 D1 \includegraphics[width=0.45\textwidth]{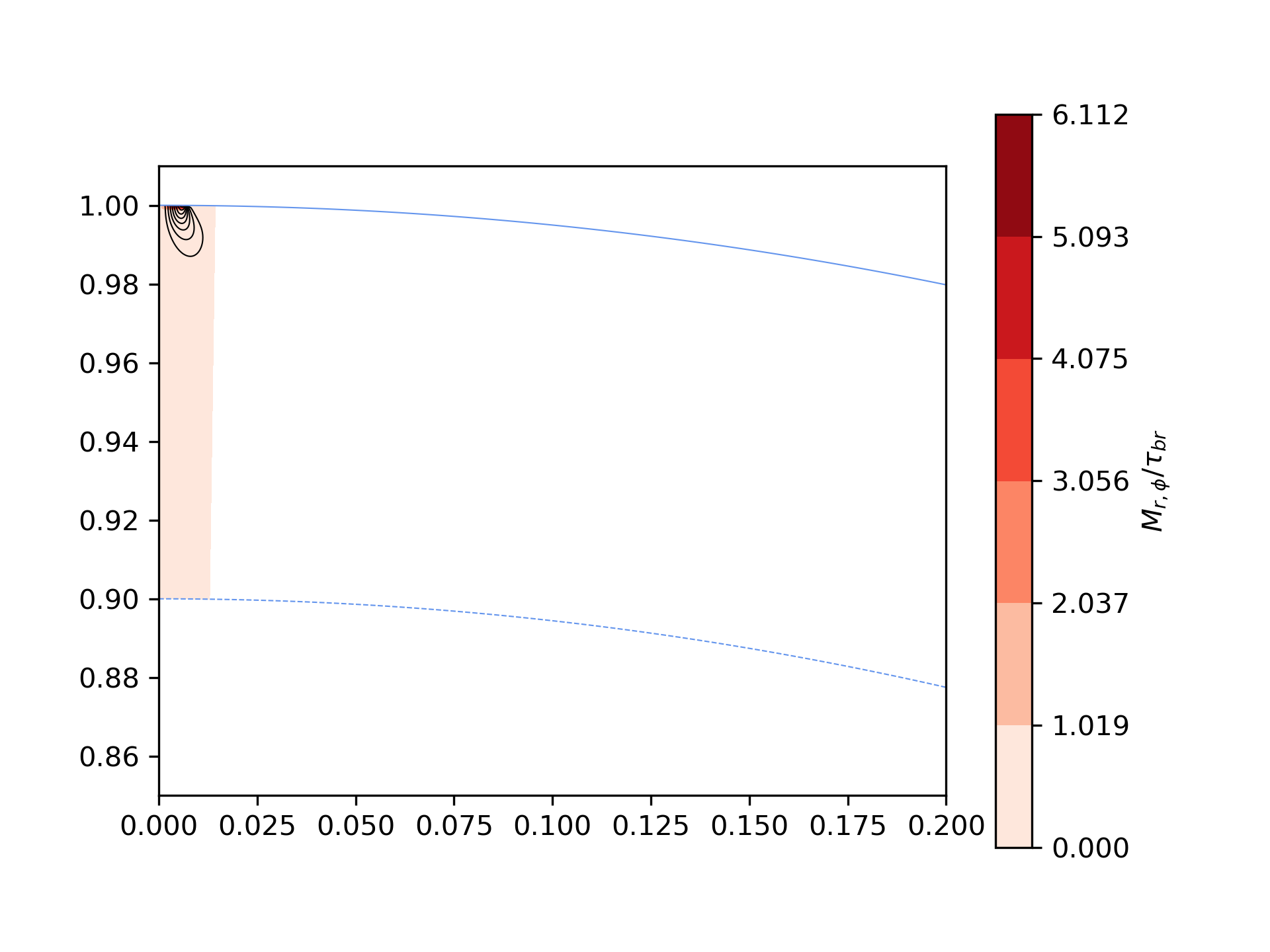}
    \caption{Plots of electric current flow-lines in black and the ratio of $M_{r\phi}/\tau_{br}$ in color for models A1, B1, C1, D1. }
 \label{Fig:Maxwell}
\end{figure*}

The distribution of the magnetospheric current of the force-free solution is such so that the current enters the star through the central part and most of the area of the polar cap and leaves the star through a narrow ring and mostly through a current sheet flowing on the separatrix between the open and closed magnetic field lines (here we have assumed that the magnetic moment and the angular momentum are parallel). Because of the singularity in the density of the current sheet, the Lorentz force and torque per unit volume become formally infinite at the edge of the polar cap. This is illustrated in Figures 2-6 by the convergence of the current flow lines at the edge of the polar cap. We note however that these are integrable singularities and the physical quantities associated to them (i.e.~torque, net force) remain finite once we integrate over the corresponding volume. The possibility of crust yielding does not depend on the local value of the force density, but rather by comparing the Maxwell stress to the the breaking stress, i.e. eqs.~(\ref{eqstr}) and (\ref{eqellim}). Indeed, Maxwell stresses remain finite and are a few orders of magnitude below $\tau_{br}$ for models A1 and C1. The crust does not yield either in model B1, but the ratio becomes $M_{r\phi}/\tau_{br}=0.2$ at the outermost layer of the integration domain. On the contrary, in model D1 the maximum shear stress becomes $M^{\rm max}_{r\phi} = 5.1 \times 10^{20}$~erg~cm$^{-3}$ close to surface, which is high enough to  cause crust yielding for a few meters ($\sim10$~m) below the surface, see Fig.~\ref{Fig:Maxwell}. Quite remarkably, the stress does not peak below the rim of the polar cap, where the current sheet enters the neutron star and the electric current density is the highest, but at some intermediate angle $\theta\approx 0.8 \theta_{pc}$. This is because the Maxwell stress is proportional to $B_{\phi}= 2I/(cr\sin\theta)$, which becomes maximum at some intermediate angle. The Maxwell stress is zero on the axis ($\theta=0$). This is because the $I(r_{rm ns},\theta)$ becomes proportional to $\sin^2 \theta$ as $\theta \to 0$ \citep{Timokhin:2006}. Thus, if there is a part of the crust more likely to yield, this will be a ring of semi-opening angle $\approx 0.8 \theta_{pc}$, rather than the region where the torque reverses from spin-down to spin-up, or even the axis. 

The possibility of crust yielding and the maximum depth where this could occur depend strongly on the detailed physics of the outer crust. As the maximum stresses appear near the conventional surface of the neutron star ($\rho \sim10^{6}$~g~cm$^{-3}$), this essentially lies at the interface between the ocean and the ion lattice. A hotter neutron star could have a deeper ocean. In practice, this implies that these stresses act on the fluid part of the crust, where eq.~(\ref{eqellim}) is no longer applicable.

\begin{figure}
    \includegraphics[width=0.5\textwidth]{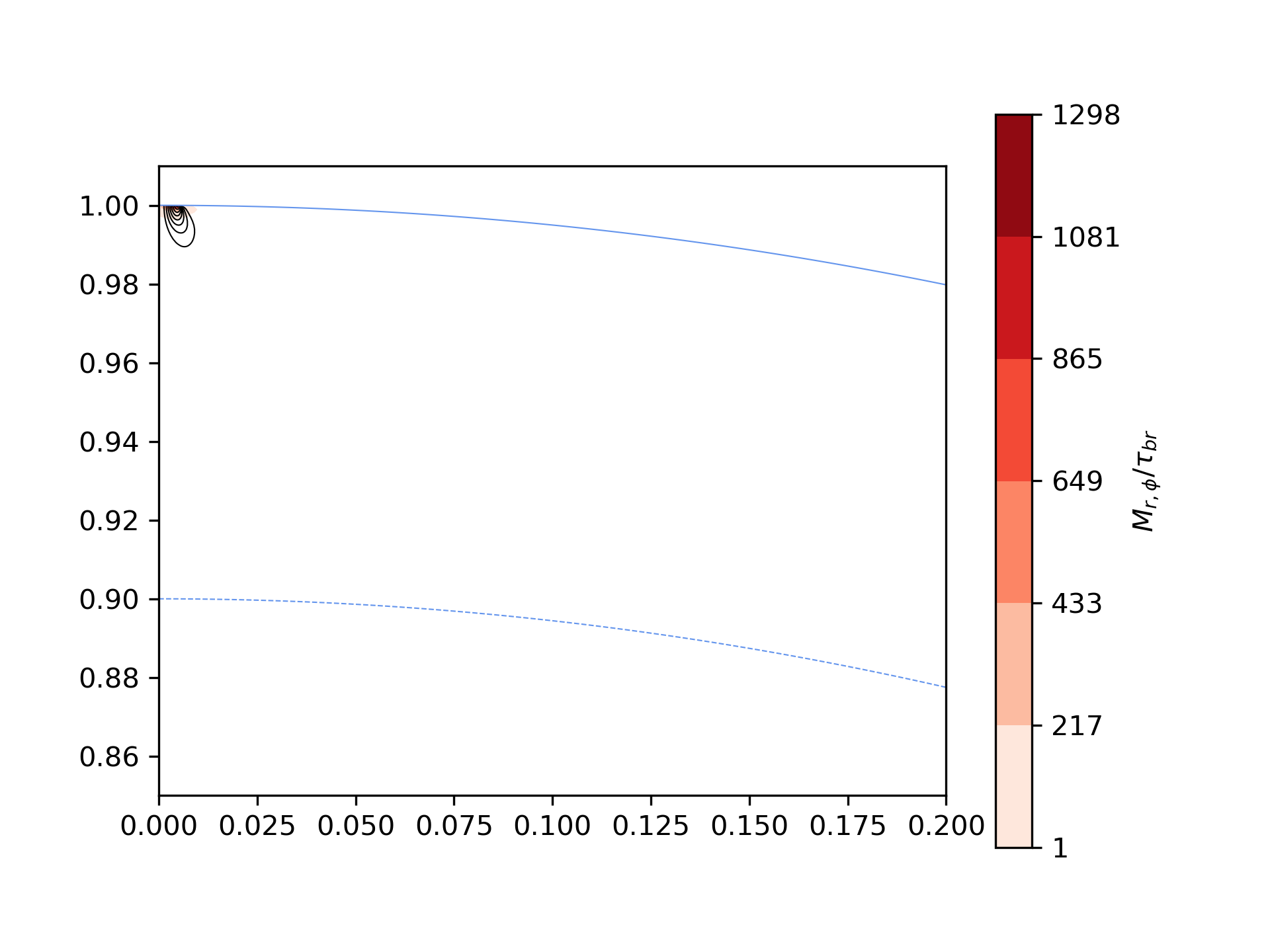}
   \includegraphics[width=0.5\textwidth]{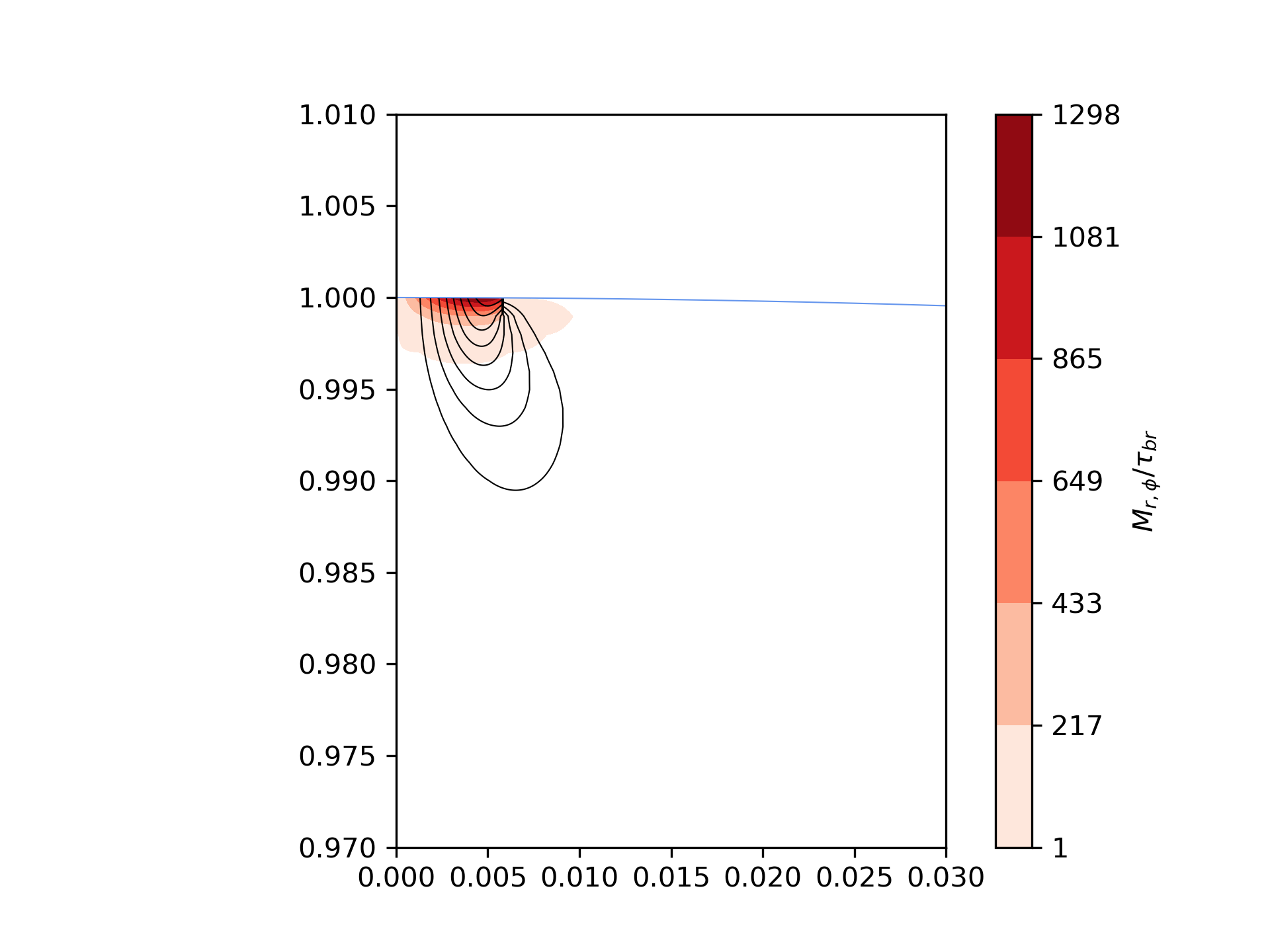}
  \caption{Magnetic stresses normalised to the breaking stress for the SGR 1806$-$20 model. Top: the entire polar cap. Bottom: zoom-in at the surface.}
  \label{figstrrsgr}
\end{figure}

Pushing the question of crust yielding to the extreme, we have also considered the magnetar with the highest known magnetic field SGR 1806$-$20 \citep{Woods:2007}, which has a long period $P=7.54$~s  and thus a very small polar cap $\theta_{\rm pc}=0.33^{\rm o}$. Its inferred dipole magnetic field is $B=2 \times 10^{15}$~G, and is the most prominent candidate  for  crust yielding. Integrating eq.~(\ref{MINIMdeq}) we find that the current reaches a depth of only 40~m beneath the surface. The spin-down torque $N_{\rm tot, ~G}=8.0\times 10^{34}$~erg is in agreement within $2\%$ with $N_{\rm align}=8.2\times 10^{34}$~erg. We find that $M^{\rm max}_{r\phi} = 1.1 \times 10^{23}$~erg~cm$^{-3}$ near the surface, which implies that the magnetospheric current will be extremely high to cause crust yielding to about 30~m below the surface (see Fig.~\ref{figstrrsgr}). Such an event will be energetically unimportant compared to the energy that could potentially be released by magnetar activity. Nevertheless, it may impact the coupling between the magnetosphere and the crust, and therefore, the spin-down efficiency. This could be related to the higher timing irregularities that are observed in strongly magnetised neutron stars and magnetars \citep{Hobbs:2010}. The timing noise in neutron stars with polar magnetic fields below $10^{12}$~G is independent of the magnetic field strength, whereas, in neutron stars with magnetic fields above this value it tends to increase and scale strongly with the magnetic field strength \citep{Tsang:2013}. It is conceivable that such behavior is related with crust yielding near the surface. Stronger magnetic fields lead to deeper crust failure. Given the episodic nature of crust failure \citep{Thompson:2017}, the loss and recovery of the spin-down current coupling with the neutron star could manifest itself as erratic variations of the spin-down, i.e. practically as timing noise.

A possible consequence of the shallow penetration of the electric current and consequently the inefficient coupling between the spin-down current and the crust, can be the lack of isolated neutron stars with long periods, with a cut-off period in the range of $24$s across the entire pulsar population \citep{Tan:2018}. This effect has been previously attributed to magnetic field decay \citep{Pons:2013}, alignment between the magnetic axis \citep{Johnston:2017} and observational selection effects \citep{Faucher:2006ApJ}. In the current picture, we note that a pulsar with a rotation period of $15$s will have a polar cap opening angle of $0.2^{\rm o}$ and the electric current will penetrate to a depth of $30$m which could lead to poor coupling and inefficient spin-down, therefore, it would be even harder for these pulsars to move to lower periods.

\subsection{Twisted magnetospheres}

In rotation powered radio pulsars, electric currents are associated with their spin-down. On the contrary, strongly magnetised neutron stars may have electric currents that support twisted magnetospheric structures \citep{Beloborodov:2009}. According to this paradigm, electric current bundles form near the surface of the star and accelerate particles that bombard the surface of magnetars, thus generating X-ray emission. To determine the crustal current that supports such structures, one can follow the internal magnetic field evolution and solve self-consistently for the crust and the magnetosphere \citep{Akgun:2018}, or alternatively consider an MHD equilibrium state taking into account the internal and external field \citep{Glampedakis:2014}. Under the approach presented in this work, one can use the minimisation technique proposed to determine the minimum crustal electric current required to generate such a bundle. Using order-of-magnitude estimates, we find that the current supporting the bundle will be very much higher than the spin-down current, scaling approximately by a factor $(r_{\rm lc}/l_{\rm b})^2$, where $l_{\rm b}$ is the size of the bundle. Assuming that the bundle is comparable to the thickness of the crust, and considering a slowly spinning magnetar, this factor could be on the order of $10^{12}$. This would thus bring the magnetic energy dissipation rate in the range of $10^{32}$~erg~s$^{-1}$. This implies that a non-neglible fraction of the bundle energy may be dissipated inside the crust.

\section{Conclusions}

In this study we have explored the closure of the magnetospheric electric current through the neutron star crust in the simplest case of axisymmetry and steady-state. We have treated the crust and the magnetosphere as a global electric circuit, where the stellar rotation generates a poloidal electric current along the ``infinitely conducting magnetic field wires'' in the magnetosphere. This is very different from previous studies which proposed that the current penetrates only within a thin surface layer in a manner similar to the interaction of electromagnetic waves with the surface charges of a perfect conductor \citep{Michel:1991, Beskin:1993, Beskin:2007}. Skin-depth penetration refers to the interaction of an externally generated electromagnetic wave with a conductor. In our case, the magnetic field already penetrates deep into the stellar interior because the latter is the source of the magnetic field, and the stellar rotation is the generator (battery) of the large scale poloidal electric current. We must acknowledge, though, that, while the axisymmetric case offers important insight on the overall properties of this current, a more interesting and complicated situation arises once the three-dimensional magnetosphere is considererd. Indeed, if the magnetic and rotation axes are not aligned, calculating the electric current distribution in the stellar interior becomes highly non-trivial.

We also remark, that the adopted dipolar form for the magnetic field, while being the norm in models of pulsar magnetospheres, it could be a simplified picture of the realistic magnetic field structure. More complex magnetic fields are likely to be present in magnetars and even older neutron stars \citep{Gourgouliatos:2018}. If this is the case, the solution for the magnetosphere and consequently the crustal electric current would become more complicated \citep{Gralla:2017}.  

We have found that the magnetospheric current, responsible for the pulsar spin-down, enters the crust and reaches its base only if we consider rapidly rotating millisecond pulsars. In the case of slower pulsars, with periods longer than $1$~s, the bulk of the electric current reaches depths less than $100$~m. Even in the case of shallow current, the crust remains within its elastic limit  without yielding, provided the magnetic field of the star is $10^{12}$~G or less. Ohmic losses in the crust are found to be orders of magnitude below the spin-down power. 

While slower spinning neutron stars (old radio pulsars close to the death line, young magnetars) have rather small polar caps and the whole magnetospheric current closes through a narrow and shallow region of the crust, we find no stresses that exceed the elastic limit, except possibly in the outer few meters of the crust (as e.g. in SGR 1806$-$20). Energetically, such an effect may be insignificant compared to the X-ray power radiated by strongly magnetised neutron stars. Nevertheless, it affects the efficiency of the coupling and generates timing noise due to torque variations.

\section*{Acknowledgements}

VK acknowledges COST ACTION PHAROS (CA16214) for STSM Grant 41713 that funded a visit to Durham University, and the Department of Mathematical Sciences of Durham University for hospitality and funding. KNG thanks Andrew Cumming and Anthony Yeates for useful discussions on the formulation of the problem and Jose Pons for suggesting the potential application to twisted magnetospheres of magnetars. We thank the referee of this paper for pointing us the correction factor of $0.94$ in eq.~(\ref{naleq}), and removing a systematic discrepancy of about $5\%$ between the torque evaluated through the integral of eq.~(\ref{itoreq}) and analytically through eq.~(\ref{naleq}).

\bibliographystyle{mnras}
\bibliography{references.bib}

\newpage
\appendix

\section{Minimisation of Ohmic losses}
\label{Appendix1}

Here we present the derivation equation \ref{MINIM} through a minimisation principle. 
Let
\begin{eqnarray}
E({\bf B})=\int_V \frac{\left(\nabla \times {\bf B}\right)^2}{\sigma} dV\,
\end{eqnarray}
and consider a variation ${\bf h}$ that vanishes at the boundary of $V$ so that  ${\bf h}|_{\partial V}={\bf 0}$, as the magnetic field ${\bf B}$ is given on the boundaries of the domain. We then define
\begin{eqnarray}
V({\bf B}, {\bf h})&=&\lim_{\epsilon\to 0}\frac{E({\bf B}
+\epsilon{\bf h })-E({\bf B})}{\epsilon}\nonumber \\
&=&2 \int_V \frac{\left(\nabla \times {\bf B} \right)\cdot \left(\nabla \times {\bf h}\right)}{\sigma} dV.
\label{minimiser}
\end{eqnarray}
$E({\bf B})$ will have a minimum, as $E>0$ if $V({\bf B}, {\bf h})=0$. 
Let us further define 
\begin{eqnarray}
{\bf A}= \frac{\nabla \times {\bf B}}{\sigma}\,,
\end{eqnarray}
and use from vector calculus the identity:
\begin{eqnarray}
\nabla \cdot \left({\bf A}\times {\bf h}\right)=-{\bf A} \cdot \nabla \times {\bf h}+{\bf h}\cdot \nabla \times {\bf A}\,.
\end{eqnarray}
Then equation \ref{minimiser} becomes:
\begin{align}
\int_V {\bf A}\cdot \nabla \times {\bf h} ~dV = \nonumber \\
\int_V {\bf h} \cdot \left(\nabla \times {\bf A}\right) dV-\int_V \nabla \cdot \left({\bf A}\times {\bf h}\right)dV=\nonumber \\
\int_V {\bf h} \cdot \left(\nabla \times {\bf A}\right) dV+\int_{\partial V} \left({\bf A}\times {\bf h}\right)\cdot d {\bf S} 
\end{align}
where we have used the divergence theorem. The second integral in the last equation is zero as ${\bf h}$ vanishes on the boundary of the domain. The first integral needs to be zero for any choice of ${\bf h}$. This is possible only if $\nabla \times {\bf A}={\bf 0}$ thus
\begin{eqnarray}
\nabla \times \left(\frac{\nabla \times {\bf B}}{\sigma}\right)={\bf 0}\,,
\end{eqnarray}
which is eq.~(\ref{MINIM}).

\label{lastpage}

\end{document}